\documentclass[%
 aip,
 amsmath,amssymb,
 reprint,%
]{revtex4-1}

\usepackage{graphicx}
\usepackage{dcolumn}
\usepackage{bm}

\usepackage[utf8]{inputenc}
\usepackage[T1]{fontenc}
\usepackage{mathptmx}
\usepackage{etoolbox}
\usepackage{bbm,bm,braket,amsmath,amssymb,amsfonts}

\def\one{\mathbbm{1}}
\def\brick{{\rm br}}
\def\IRF{{\rm IRF}}
\def\kket#1{{| #1\rangle\!\rangle}}
\def\bbra#1{{\langle\!\langle #1|}}
\def\bbraket#1{{\langle\!\langle #1 \rangle\!\rangle}}
\def\One{\mathbb{I}}

\makeatletter
\def\@email#1#2{%
 \endgroup
 \patchcmd{\titleblock@produce}
  {\frontmatter@RRAPformat}
  {\frontmatter@RRAPformat{\produce@RRAP{*#1\href{mailto:#2}{#2}}}\frontmatter@RRAPformat}
  {}{}
}%
\makeatother
\begin{document}


\title[Chaos Round--a--Face]{Many Body Quantum Chaos and Dual Unitarity Round--a--Face}

\author{Toma\v z Prosen}

 \affiliation{Faculty of Mathematics and Physics, Universty of Ljubljana, Jadranska 19, Si-1000 Ljubljana, Slovenia}
 
 \email{tomaz.prosen@fmf.uni-lj.si}

\date{\today}

\begin{abstract}
We propose a new type of locally interacting quantum circuits which are generated by {\em unitary} interactions round--a--face (IRF). Specifically, we discuss a set (or manifold) of {\em dual-unitary} IRFs with local Hilbert space dimension $d$ (DUIRF$(d)$) which generate unitary evolutions both in space and time directions of an extended 1+1 dimensional lattice. We show how arbitrary dynamical correlation functions of local observables can be evaluated in terms of finite dimensional completely positive trace preserving unital maps, in complete analogy to recently studied circuits made of dual unitary brick gates (DUBG). In fact, we show that the simplest non-trivial (non-vanishing) local correlation functions in dual-unitary IRF circuits involve observables non-trivially supported on at least two sites.
We completely characterise the 10-dimensional manifold of DUIRF$(2)$ for qubits ($d=2$) and provide, for $d=3,4,5,6,7$, empirical estimates of its dimensionality based on numerically determined dimensions of tangent spaces at an ensemble of random instances of dual-unitary IRF gates. In parallel, we apply the same algorithm to determine ${\rm dim}\,{\rm DUBG}(d)$ and show that they are of similar order though systematically larger than ${\rm dim}\,{\rm DUIRF}(d)$ for $d=2,3,4,5,6,7$. It is remarkable that both sets have rather complex topology for $d\ge 3$ in the sense that the dimension of the tangent space varies among different randomly generated points of the set. Finally, we provide additional data on dimensionality of the chiral extension of DUBG circuits with distinct local Hilbert spaces of dimensions $d\neq d'$ residing at even/odd lattice sites.
\end{abstract}

\maketitle

\begin{quotation}
Quantum many-body dynamics of generic interacting systems is essentially intractable and is amenable only to quantum simulation.
One may wonder, whether there exist
non-integrable (generically, quantum chaotic) many-body systems with local interactions which would have exactly solvable spatio-temporal correlation functions of local observables. 
These models would be understood as quantum many-body analogs of baker and cat maps, playing a similar role in classical single-particle chaos.
We outline two complementary classes of quantum dynamical systems with exactly solvable dynamical correlations exhibiting a rich ergodic hierarchy of dynamical behaviors: the dual-unitary brickwork circuits and, newly proposed dual-unitary face models (circuits with dual-unitary interactions round--a--face (DUIRF)). Remarkably, dynamical correlation functions of local observables in these families of 1+1 dimensional interacting systems are non-vanishing only along the edges of causal cones, where they are given in terms of dissipative single-particle quantum (Markov) dynamical systems. The latter in turn can be clearly classified as non-ergodic, ergodic and mixing, based on the spectrum of finite-dimensional quantum Markov matrix.
The dynamical and geometric features of such DUIRF dynamical systems are discussed in relation to previously studied dual unitary circuits. We conjecture that recent exact results on random matrix spectral statistics, entanglement dynamics and operator spreading in dual-unitary brickwork circuits can be adapted to dual-unitary IRF circuits.
\end{quotation}

\section{\label{sec:intro}Introduction}

Precise definition of quantum chaos of many-body quantum systems has been elusive for a long time, even in the simplest context of quantum spin-lattice systems with local interactions. For example,
it has been observed a while ago\cite{Mila1993,Poilblanc1993,Hsu1993} that spectral statistics of non-integrable spin-1/2 chain Hamiltonians with nearest-neighbor interactions conform to Random matrix theory\cite{Mehta} (RMT) and the match to RMT statistics on appropriate physical time/energy scales has been considered as a working definition of quantum chaos in condensed matter theory community for decades. Nevertheless, the first analytical 
explanations\cite{KLP18,ChalkerPRX,ChalkerPRL,Friedman,Roy} or 
proofs\cite{BKP18,BKP20} of this \emph{quantum chaos conjecture} came only very recently, and only in quite restricted contexts. On the other front, people have been trying to identify the quantum analogs of Lyapunov exponents\cite{Maldacena,Swingle} and Kolmogorov-Sinai (dynamical) entropies (characterising algorithmic complexity of dynamics)\cite{Yoshida}. Such quantum dynamical entropies\cite{Alicki}, however, cannot even discriminate between free and interacting evolutions in the thermodynamic limit, as the information (entropy) is generically propagating from an (infinite) bath of degrees of freedom to the subsystem of interest \cite{TP07}, and hence obscuring `dynamical generation' of entropy (the notion which thus cannot be precisely defined in extended systems). On the other hand, quantum Lyapunov exponents have been nevertheless defined through the out-of-time-ordered correlation functions (OTOCs), but these construction is meaningful only in the so-called ``large $N$'' theories (models) which are essentially semiclassical (with an effective Planck constant $\hbar=1/N$).  Thus, the genuinely hardest, and arguably the most interesting cases are related to understanding of dynamical complexity in spin lattice models with finite local Hilbert space dimension and with local interactions. The simplest among those are the local quantum circuit models, which can be understood as a discrete-time quantum dynamical systems on a lattice, or quantum cellular automata\cite{Farrelly,QCA}.

Recently, a substantial progress has been achieved in understanding quantum chaos conjecture and dynamical complexity in (Floquet) local quantum circuit models\cite{ChalkerPRX,nahum2017quantum,nahum2018operator,khemani,vonKeyserlingk2018operator} In those models, besides spatio-temporal OTOCs the most fruitful measure of dynamical complexity has been identified as the operator-space entanglement entropy \cite{PP07}. The latter quantifies the so-called operator spreading, or growing bipartite correlations of time-dependent local operators interpreted as elements of tensor products of local Hilbert spaces. Moreover, it has been shown that explicit and exact results on RMT spectral correlations\cite{BKP18,BKP20}, dynamical correlation functions\cite{bertini2019exact,austen_prl},
quantum quenches\cite{Lorenzo}, (operator) entanglement dynamics\cite{PRX19,bertini2020operatorI,bertini2020operatorII,Reid}, information scrambling\cite{BrunoLorenzo}, and OTOCs\cite{austen_PRR}, can be obtained even for local qudit circuits (with fixed local Hilbert space dimension $d$, say $d=2$) provided the circuit, i.e. the local gates, satisfy the so-called dual-unitarity (DU) condition\cite{bertini2019exact}. It has been shown that DU circuits include integrable and (generically) non-integrable (chaotic) systems\cite{bertini2019exact}, in particular the previously studied self-dual kicked Ising model\cite{Sarang}. Studying {\em space-time duality} proved useful also to get important new insights into the behavior of non-DU circuits\cite{Gutkin_JPA,Gutkin_pert,Amos,Garratt,Ippoliti1,Ippoliti2,Grover,Abanin,Pavel}. 

DU circuits are thus a representative class of exactly solvable chaotic quantum systems, very much like the baker and cat maps in classical chaos theory\cite{ott}. In analogy to structural stability of hyperbolic flows \cite{Robbin,Robinson} in classical chaos theory we conjectured (and found partial evidence of) \cite{PRX2021} perturbative stability of DU quantum dynamical systems.

In this paper we propose an extension of a class of local quantum circuits in terms of a concept of \emph{unitary} interactions round--a--face (IRF). Unitary IRF circuits can be thought of as a complementary model to brickwork quantum circuits and yet another realization of quantum cellular automata. Specifically, IRF gate is just a controlled (or kinetically constrained) local unitary gate, where the control is placed on the neigbouring two qudits and could hence capture the dynamics of (Floquet) driven Rydberg atom chains\cite{Lukin} or similar manipulated systems. As a deterministic version of unitary IRF dynamics, we should mention a rule 54 reversible cellular automaton \cite{rule54review}. While Yang-Baxter integrable IRF models (also known as RSOS models)\cite{baxter,forrester} can give rise to integrable quantum spin chain Hamiltonians\cite{Pearce}, it is not clear if unitary integrable IRF circuits can be generated beyond the singular case of classical reversible cellular automata mentioned earlier\cite{bobenko1993two,rule54review} (for which Yang-Baxter structure is not clear at the moment anyway). Another related integrable kinetically constrained continuous time (Hamiltonian) dynamics has been studied in Ref.~\cite{Lenart1,Lenart2,Vernier}.

We then extend the concept of IRF circuits to DU IRF circuits of qudits ($d=2,3\ldots$). We show that, similarly as for DU brickwork circuits, the space-time correlation functions of any local observable supported on a pair of neighbouring sites can be shown to be non-vanishing only along two light-rays, where it is evaluated in terms of a pair of completely positive, trace preserving, unital maps acting on pairs of qudits (note that for DU brickwork circuits the correspondig maps act on a single qudit). This map can in fact be interpreted as a classical Markov chain as it acts non-trivially only on a $(d-1)^2 + 1$ dimensional subspace spanned by diagonal operators with vanishing partial traces plus the identity operator. 

We show how to completely characterise DUIRF circuits of qubits, $d=2$, and explicitly parametrize the corresponding 10-dimensional manifold ${\rm DUIRF}(2)$. We also empirically estimate dimensions of ${\rm DUIRF}(d)$ and of related DU brick gates ${\rm DUBG}(d,d')$ (where dimensions of local Hilbert spaces on even and odd checkerboard sublattices of the brickwork, $d$ and $d'$ respectively, can be different), for $d (d') = 3,4,5,6,7$.
It is remarkable that both sets ${\rm DUIRF}(d)$, ${\rm DUBG}(d,d')$ have non-uniform dimensions, i.e. the dimensions of tangent space at different generic (random) elements of the set are different for $d,d'\ge 3$.
Nevertheless, we find consistently that
${\rm dim}\,{\rm DUIRF}(d) > {\rm dim}\,{\rm DUBG}(d,d)$, locally everywhere, i.e. for all elements of the sets. We sketch as well some other interesting problems that one could approach using DUIRF circuits, most specifically the problem of spectral statistics and the idea of the proof of RMT spectral form factor for DU IRF circuits.

\section{Unitary IRF circuits}

\begin{figure}
\includegraphics[scale=0.45]{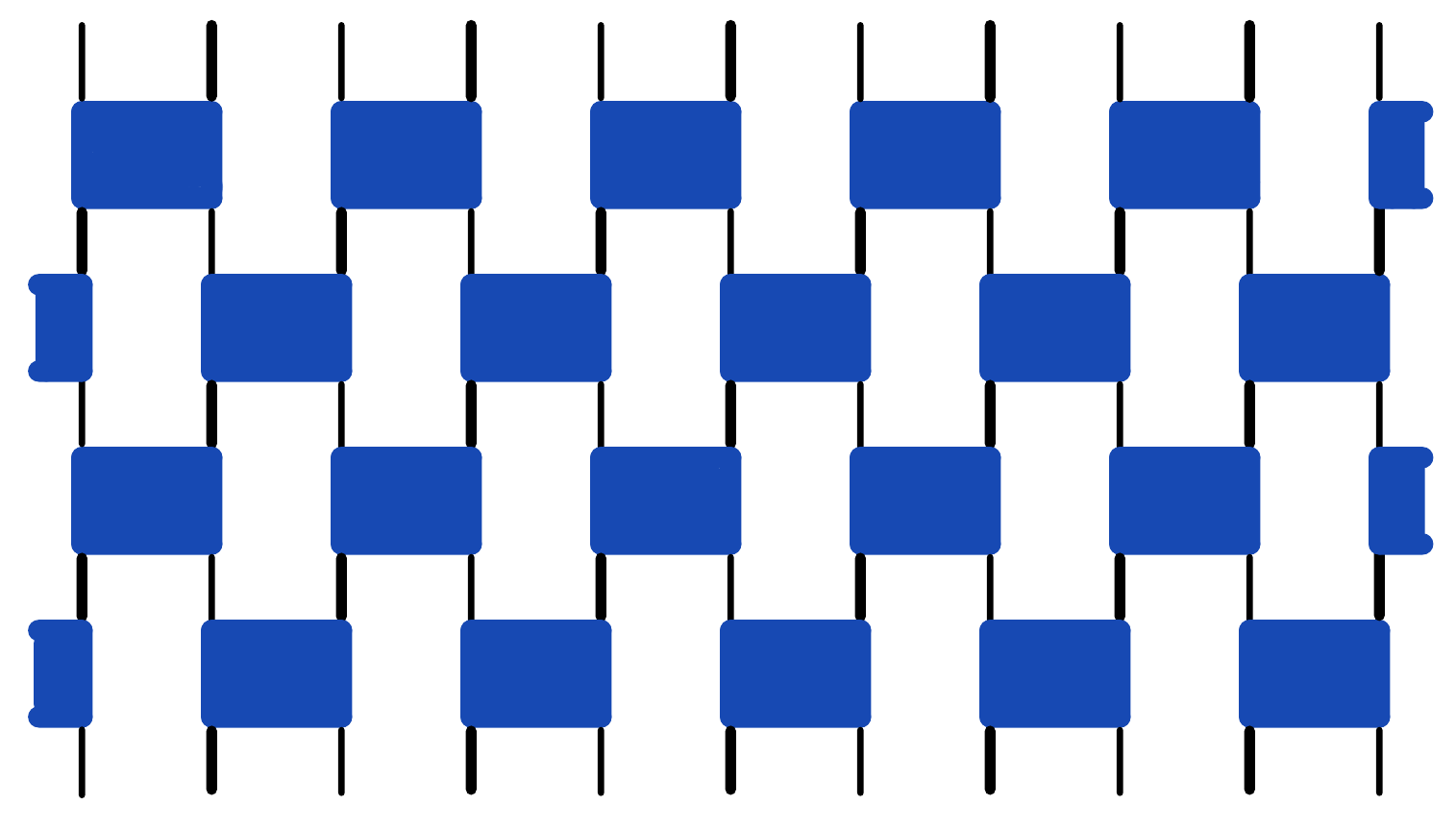}
\hspace{1mm}
\includegraphics[scale=0.6]{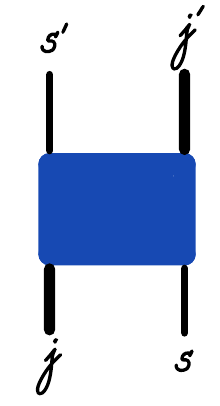}
\caption{\label{fig:BWcircuit} Brickwork local circuit composed of brick gates (local gate indicated on the right) for $t=2$ (depth 4). Note that the dimensions of the even/odd local spaces could be different (indicated by thin/thick wires). Evolution time runs bottom-up throughout the paper.}
\end{figure}

Let us consider a chain of even number, $2L$, $L\in\mathbb{N}$, of qudits ($d$-level quantum systems), such that the Hilbert space of the system is given as a $d^{2L}$ dimensional tensor product
${\cal H} = {\cal H}_1^{\otimes 2L}$, ${\cal H}_1 = \mathbb{C}^{d}$.

We may also consider a more general, \emph{chiral} situation, where the pair of neighboring sites have different Hilbert space dimensions
${\cal H}_1 = \mathbb{C}^d$, ${\cal H}'_1 = \mathbb{C}^{d'}$, and two isomorphic system Hilbert spaces (with even/odd sublattices interchanged),
${\cal H} = ({\cal H}_1\otimes {\cal H}'_1)^{\otimes L}$,
${\cal H}'= ({\cal H}'_1\otimes {\cal H}_1)^{\otimes L}$.
In many physical situations, such as when discussing periodically driven (Floquet) spin chains, or Trotterized Hamiltonian evolutions with local one-dimensional interaction (in the latter only the case $d'=d$ makes sense), as well as in protocols for analog quantum simulation\cite{Nori} of local interactions, it is customary to consider brickwork quantum circuits. For simplicity, we assume space-time homogeneity\footnote{The extension of the entire analysis in our paper to arbitrarily spatially and temporally modulated gates is straightforward, so we do not mention it any further}. Hence, considering a single unitary gate $U \in {\rm U}(d d')$ interpreted as a linear map
${\cal H}_1\otimes {\cal H}'_1 \to
{\cal H}'_1\otimes {\cal H}_1$, or in explicit matrix/Dirac notation\footnote{Throughout the paper we use the convention that row/column of a matrix is labeled by a lower/upper index, while multi-indices at the same level label tensor- (Kronecker-) product spaces.}
\begin{equation}
    U^\brick=\sum_{j,j'=1}^d \sum_{s,s'=1}^{d'}
    U_{j\,s}^{s'j'} \ket{s'}\otimes \ket{j'} \bra{j} \otimes \bra{s}\,,
    \label{eq:BG}
\end{equation}
we define a generator (or Floquet propagator) of a brickwork local circuit as 
\begin{equation}
\mathcal U = 
\mathcal U^{\rm o} \mathcal U^{\rm e} \;:\; {\cal H} \to {\cal H}
\label{eq:UF}
\end{equation} where
\begin{eqnarray}
\mathcal U^{\rm e} &=& \prod_{x=1}^L U^\brick_{2x-1,2x} \;:\; {\cal H} \to {\cal H}',\label{eq:BWC}\\
\mathcal U^{\rm o} &=& \prod_{x=1}^L U^{\brick}_{2x,2x+1} \;:\; {\cal H}' \to {\cal H}',\nonumber
\end{eqnarray}
and where the subscripts in $U^\brick_{x,y}$ denote the positions $x,y$ of two qudits (sites) where the brick gate $U^\brick$ acts non-trivially (see Fig.~\ref{fig:BWcircuit} for an unambiguous graphical definition). Periodic boundaries are assumed throughout: $x+2L\equiv x$. 

Although we will use brickwork circuits later for comparison, we make a twist in this paper and propose to study another physics paradigm of \emph{generic local spatiotemporal dynamics on 1+1 dimensional lattice}. Specifically, we propose unitary face circuits where the local interactions are given in terms of nearest-neighbor controlled (e.g., kinetically constrained) local unitary gates or, equivalently, in terms of unitary interactions round--a--face.
Here we assume all local spaces to be isomorphic\footnote{We could as well define more general class of IRF circuits where dimensions of local spaces at even/odd lattice sites would be different, $d\neq d'$, but this would require two different IRF gates at even/odd half-time steps (\ref{eq:BWC}) and will not be considered here.} $d=d'$.

\begin{figure}
\includegraphics[scale=0.57]{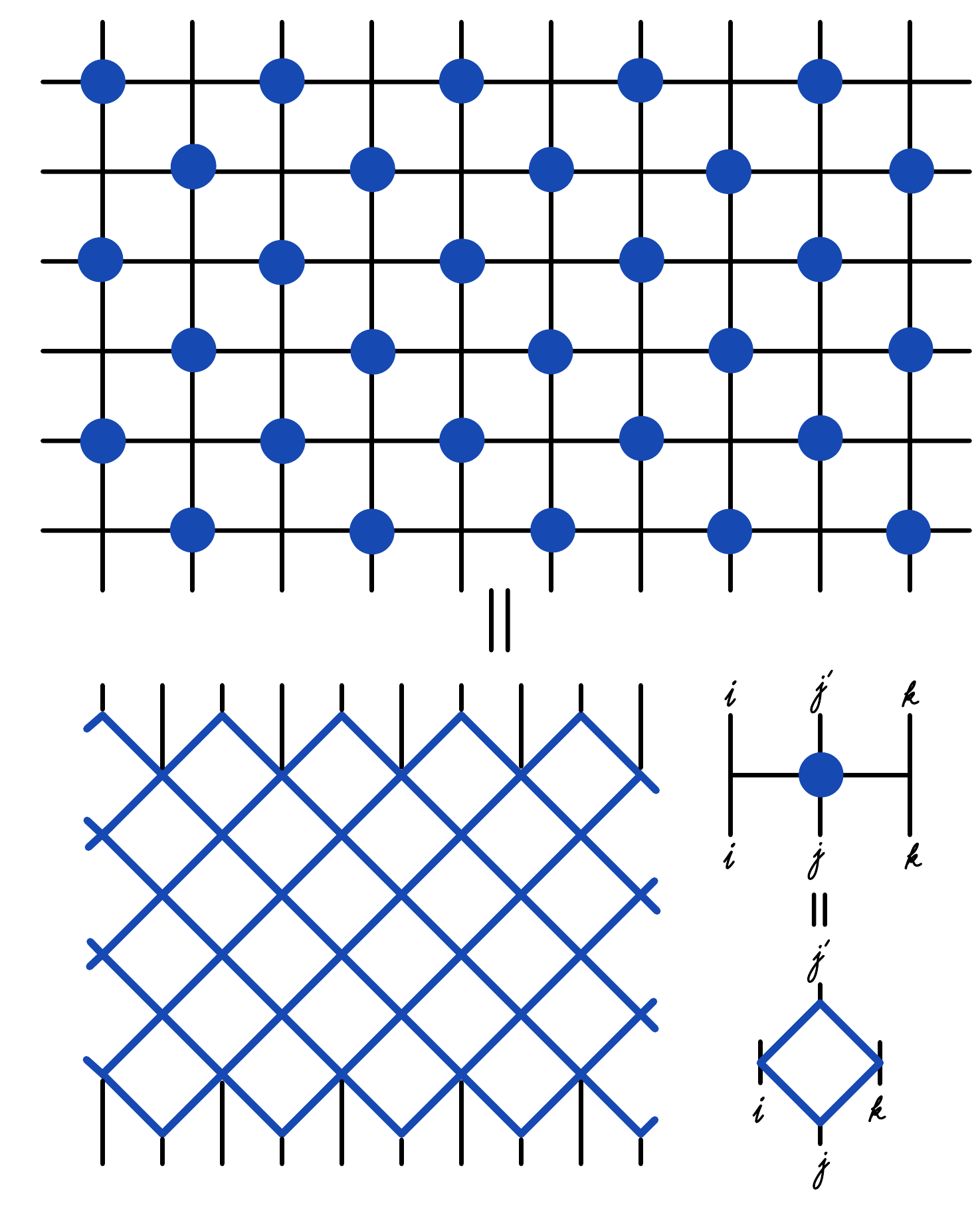}
\caption{\label{fig:IRFcircuit} 
Face local circuit composed of IRF gates
(local gate indicated on the bottom-right) with duration $t=3$ (depth 6), using two different notations, either in terms of controlled unitary gates (top) or face plaquettes (bottom).}
\end{figure}

Consider a set of $d^2$ arbitrary unitary matrices $\{u_{ik}\in {\rm U}(d)\}_{i,k\in\{1,\ldots,d\}}$ which define a general 2-controlled 3-qudit unitary gate (as a unitary over ${\cal H}_1^{\otimes 3}$)
\begin{equation}
    U^\IRF = \sum_{i,j,k,j'=1}^d (u_{ik})_j^{j'} \ket{i}\otimes \ket{j'}\otimes \ket{k} \bra{i}\otimes\bra{j}\otimes\bra{k}\,
    \label{eq:UIRF}
\end{equation}
Equivalently, a set of $d^4$ amplitudes $(u_{ik})_{j}^{j'}$ can be understood as defining a (unitary) IRF model (see Fig.~\ref{fig:IRFcircuit}).
Such 3-qudit gates, embedded into the many-body Hilbert space ${\cal H}$ as $U^\IRF_{x-1,x,x+1}$ now define locally interacting unitary circuit with the generator of the form (\ref{eq:UF}), where
\begin{eqnarray}
\mathcal U^{\rm e} &=& \prod_{x=1}^L U^\IRF_{2x-1,2x,2x+1}, \label{eq:IRFC}\\
\mathcal U^{\rm o} &=& \prod_{x=1}^L U^\IRF_{2x,2x+1,2x+2}. \nonumber
\end{eqnarray}
Similarly to brickwork circuits (\ref{eq:BWC}), which behave as quantum cellular automata\cite{Farrelly}, namely they propagate information/correlation by one-site per layer of the gates, one notes the same feature for IRF circuits (\ref{eq:IRFC}).

An example of a unitary IRF circuit is a Trotterization\cite{FloquetPXP1,FloquetPXP2} of the so-called PXP model\cite{PXP,abanin_NP2018} beautifully modelling kinetically constrained Rydberg atom chains\cite{Lukin}. Specifically, the three site Hamiltonian of the PXP model
$h_{x-1,x,x+1}= P_{x-1} X_x P_{x+1}$, where 
$$P=\begin{pmatrix} 1 & 0 \cr 0 & 0\end{pmatrix}\,\quad X=\begin{pmatrix} 0 & 1 \cr 1 & 0\end{pmatrix}\,,$$
clearly exponentiates to a unitary IRF gate $U^\IRF_{x-1,x,x+1} = \exp(-{\rm i}\Delta t h_{x-1,x,x+1})$, where $\Delta t$ is the time step.
Other recently studied examples of unitary IRF cicruits are classical reversible cellular automata\cite{bobenko1993two}, like the rule 54\cite{prosen2016integrability,rule54review,katja_PRL} or the rule 201 (`classical PXP')\cite{wilkinson2020exact}.
More broadly, unitary IRF circuits represent a natural language to describe Floquet or driven quantum kinetically constrained models.

 It is interesting to note that both manifolds of brick and IRF local gates share the same number of independent real parameters (for $d'=d$), specifically $d^4$, i.e. the number of parameters of ${\rm U}(d^2)$ or the number of parameters for $d^2$ independent elements of ${\rm U}(d)$, respectively. However, we should then also mention different gauge-invariance groups of these parametrizations. While the brick gate can be transformed as
\begin{equation}
U^\brick \leftarrow (h^\dagger \otimes g^\dagger) U^\brick (g\otimes h),
\end{equation}
for arbitrary $g\in{\rm SU}(d), h\in{\rm SU}(d')$,  to yield an equivalent circuit, the IRF gate can be gauge-transformed
as
\begin{equation}
U^\IRF \leftarrow
 (\Delta^\dagger \otimes g^\dagger \otimes \Delta^\dagger)
U^\IRF (\Delta \otimes g \otimes \Delta),
\end{equation}
where $g\in{\rm SU}(d)$ arbitrary and $\Delta_{j}^{j'} = \delta_{j,j'}
e^{{\rm i}\theta_j}$,  $\theta_j \in[0,2\pi)$, is a \emph{diagonal} phase matrix (where one of the phases $\theta_j$ can be fixed without loss of generality). We thus have the following gauge groups for the two classes of circuits
\begin{eqnarray}
G^\brick{}&=&{\rm SU}(d)\otimes {\rm SU}(d'),\qquad 
\textrm{ for\;brickwork\;circuits},\\
G^\IRF&=&{\rm SU}(d)\otimes {\rm U}(1)^{\otimes (d-1)},\qquad
\textrm{ for\;IRF\;circuits}.
\label{gaugeIRF}
\end{eqnarray}

One may wish to investigate dynamics, entanglement propagation and operator spreading in IRF circuits and compare to existing results for brickwork circuits. Specifically, it would be desirable to derive analogous results to\cite{nahum2017quantum,nahum2018operator,khemani,vonKeyserlingk2018operator} for \emph{random} IRF circuits where matrices $u_{ik}$ are independent Haar-random ${\rm U}(d)$ matrices for all pairs of components $i,k$ and for each space time point. 
In this paper, however, we aim at investigating IRF circuits with an additional structure, namely, the \emph{dual-unitarity}. 

\section{Correlation decay in dual-unitary quantum lattice dynamical systems}

\subsection{Spatio-temporal correlation function and folded circuit representation}

Here we set the fundamental problem of quantum dynamics on a space-time lattice, specifically, the computation of space-time correlation function of local observables in the tracial (infinite temperature/maximum entropy) state.
Considering a pair of local traceless observables $a,b$, with $a_x,b_x$ being their embedding into ${\cal H}$ at site $x$, we aim at calculating
\begin{equation}
    C_{a,b}(x,y;t) = \lim_{L\to\infty} \frac{1}{{\rm dim}{\cal H}} {\rm tr}(a_x \mathcal U^t b_y \mathcal U^{-t}).
\end{equation}
Explicit, exact or analytical computation of correlation functions, being the fundamental importance in diverse areas of
condensed matter and statistical physics, represent an insurmountable obstacle even in the simplest (say integrable) interacting theories. Nevertheless, we will show below how the correlations can be explicitly treated in a class of generically non-integrable cuircuit models.

\begin{figure}
\includegraphics[scale=0.6]{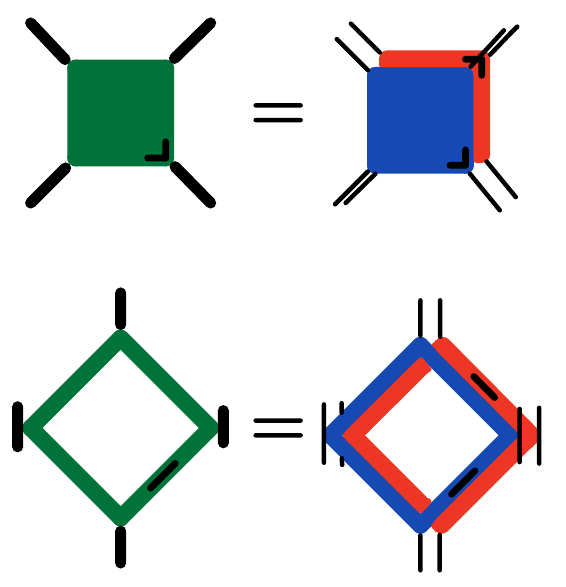}
\caption{\label{fig:5} 
Definition of the folded (Heisenberg picture) brick (top) and IRF (bottom) gate. Note that thick wires correspond to doubled Hilbert space (ket$=$left, bra$=$right thin wire).}
\end{figure}

In the so-called folded-circuit representation\cite{Banuls}, one defines a doubled (operator) Hilbert space 
$\mathcal H^{\rm op} = \mathcal H\otimes \mathcal H $, which can be considered as composed of doubled local spaces
$\mathcal H^{\rm op}_1 = \mathcal H_1\otimes \mathcal H_1 \simeq \mathbb C^{d^2}$, and possibly different local operator space $\mathcal H^{\rm op'}_1 = \mathcal H'_1\otimes \mathcal H'_1 \simeq \mathbb C^{{d'}^2}$ for even-site sublattice.
Defining doubled local brick gate over $({\mathcal H}^{\rm op}_1)^{\otimes 2}$ (Fig.~\ref{fig:5}-top)
\begin{equation}
W^\brick = U^\brick \otimes (U^\brick)^T\,
\end{equation}
where $^T$ denotes the matrix transposition, and local operator states 
\begin{eqnarray}
\kket{a} &=& \frac{1}{\sqrt{d}}\sum_{i,j}a_{i}^{j} \ket{i}\otimes\ket{j},\\ 
\kket{b} &=& \frac{1}{\sqrt{d}}\sum_{i,j} b_{i}^{j} \ket{i}\otimes\ket{j},\\
\kket{\circ} &=& \frac{1}{\sqrt{d}} \sum_{i} \ket{i}\otimes\ket{i},
\end{eqnarray}
with possibly $d$ replaced by $d'$ for even-labelled sites, 
one immediately writes an equivalent expression for the correlation function
\begin{equation}
    C_{a,b}(x,y;t) = \lim_{L\to\infty} \bbra{b_y} \mathcal W^t \kket{a_x},
    \label{eq:Cab}
\end{equation}
where $\kket{a_x} = \kket{\circ}^{\otimes (x-1)}\otimes\kket{a}\otimes\kket{\circ}^{\otimes(L-x)} $ and
$\bbra{b_y} =\bbra{\circ}^{\otimes(y-1)}\otimes\bbra{b}\otimes\bbra{\circ}^{\otimes(L-y)}$. Here $\mathcal W$ is the 
\emph{operator circuit} over $\mathcal H^{\rm op}$ built as in Eqs.(\ref{eq:UF},\ref{eq:BWC}) with $U$'s replaced by $W$'s.

Completely analogous folded circuit construction applies also for IRF circuits, where the (folded) IRF operator gate reads as (Fig.~\ref{fig:5}-bottom)
\begin{equation}
    W^\IRF = U^\IRF \otimes (U^\IRF)^T
    \label{eq:IRFfold}
\end{equation}
which is a unitary IRF gate as well (over local Hilbert spaces of dimension $d^2$).
Unitarity conditions for, respectively, brick and IRF local gates can be now expressed 
as \emph{unitality} (schematically in Figs.~\ref{fig:6},\ref{fig:7}-top) \begin{eqnarray}
W^\brick \kket{\circ}\otimes\kket{\circ} &=& \kket{\circ}\otimes\kket{\circ}, \label{eq:B1}\\
 \bbra{\circ}\otimes\bbra{\circ} W^\brick&=& \bbra{\circ}\otimes\bbra{\circ},\label{eq:B2}\\
W^\IRF \kket{\circ}\otimes\kket{\circ}
\otimes\kket{\circ}&=&\kket{\circ}\otimes\kket{\circ}\otimes\kket{\circ},\label{eq:I1}\\
 \bbra{\circ}\otimes\bbra{\circ}
\otimes\bbra{\circ}W^\IRF&=&\bbra{\circ}\otimes\bbra{\circ}\otimes\bbra{\circ}.\label{eq:I2}
\end{eqnarray}
These rules, and the fact that the operators are traceless, i.e.
$\bbraket{\circ|a}=\bbraket{\circ|b}=0$, immediately imply strict causality of the correlator, namely that the maximal speed of information propagation equals 1 (one site per circuit layer): $C_{a,b}(x,y;t)=0$ for $|x-y|>2t$. Aside from that, the computation of the correlator $C_{a,b}(x,y;t)$ for a generic local gate circuit is believed to be hard, i.e. to have a positive Kolmogorov algorithmic complexity in $t$.

\subsection{Dual-unitary brickwork circuits: review}

\begin{figure}
\includegraphics[scale=0.55]{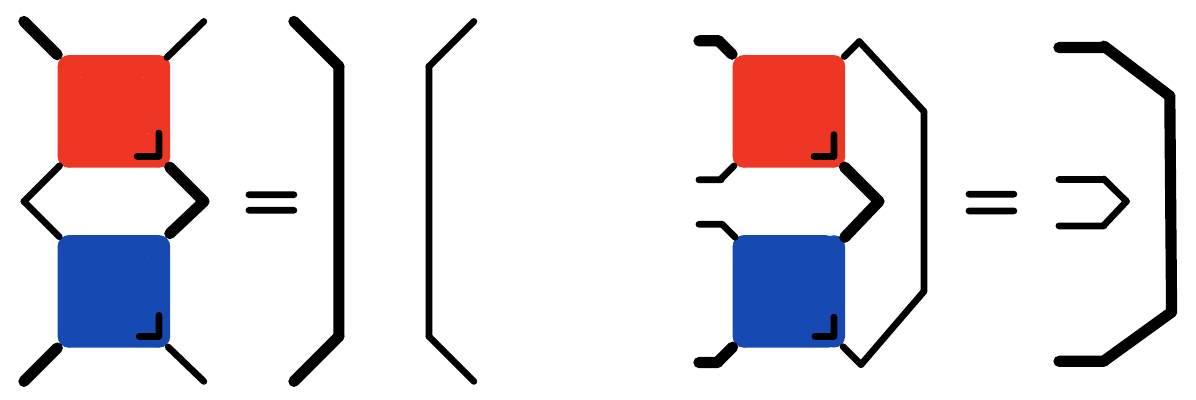}
\caption{\label{fig:3} 
Dual-unitarity: time unitarity (left) and space unitarity (right) condition for the dual-unitary brick gate (element of ${\rm DUBG}(d,d')$). Wires are drawn at $45^\circ$ angles from the gates to stress the space-time symmetry.}
\end{figure}

It has been noted in Ref.~\cite{bertini2019exact} that there exist a rich class of brickwork unitary circuits where computation of arbitrary local correlations can be drastically simplified. These are the so-called dual-unitary brickwork circuits which generate unitary dynamics not only in time (vertical) direction, but also in space (horizontal) direction. In other words, not only the local brick gate $U^\brick$ (\ref{eq:BG}) is unitary, but also the space-time reshuffled gate
\begin{equation}
\tilde{U}^\brick =\sum_{j,j'=1}^d \sum_{s,s'=1}^{d'}
    U_{j\,s'}^{s j'} \ket{s'}\otimes \ket{j'} \bra{j} \otimes \bra{s}\,,
\end{equation}
is unitary
\begin{equation}
\tilde{U}^\brick (\tilde{U}^\brick)^\dagger =\one\,.
\label{eq:DU}
\end{equation}
The gate $\tilde{U}^\brick$ is referred to as the space-time dual of $U^\brick$, and the condition (\ref{eq:DU}) (see Fig.~\ref{fig:3}-right) as space unitarity.
The gates which are both, time unitary and space unitary, form a local submanifold (locally smooth subset) ${\rm DUBG}(d,d')$ of the Lie group ${\rm U}(d d')$ and can be completely characterized \cite{bertini2019exact}
for qubits.
Specifically, one can write an arbitrary dual unitary gate
for $d=d'=2$ as
\begin{equation}
{\rm DUBG}(2,2) = \{
(u\otimes v)S e^{{\rm i}( 
\beta\one+ \gamma\,\sigma\otimes\sigma)}
(w\otimes r)\},
\end{equation}
where $u,v,w,r\in {\rm SU}(2)$, $\beta,\gamma\in\mathbb R$,
and $\sigma_j^{j'}= (-1)^{j-1} \delta_{j,j'}$ (Pauli-Z matrix),
$S_{i\,j}^{i'j'}\!=\delta_{i,j'}\delta_{j,i'}$
(SWAP, $S\ket{j}\otimes\ket{s} = \ket{s}\otimes \ket{j}$).
Counting the number of independent real parameters, one should note that out of 3 parameters (e.g. Euler angles) determining each local ${\rm SU}(2)$ gate, two can be removed, as Euler rotations around $z-$axis commute with the Ising interaction, so one is left with ${\rm dim}\,{\rm DUBG}(2,2)=12$ independent parameters.

Although large multi-parametric families of DU gates have been proposed\cite{Gutkin_Hadamard,austen_prl,balazs_private} for $d>2$, the complete characterization of ${\rm DUBG}(d,d')$ remains a challenging open problem (see section~\ref{estimates} for some empirical observations). One should note that dual-unitarity condition is equivalent to requiring that $(S U^\brick)^{T_1}$ is unitary, where $^{T_1}$ is a partial transposition. Using the result (Theorem 3.1) of Ref.~\cite{pellegrini} one can show that ${\rm DUBG}(d,d')$ can be identified with the set of {\em unital channels} over $\mathbb C^d\otimes \mathbb C^{d'}$, whose complete characterisation is, however, still open. We note that the \emph{entangling power} of such bi-partite partial-transpose unitaries have been discussed also in Refs.~\cite{PRA07,Karol}.

\begin{figure}
\includegraphics[scale=0.65]{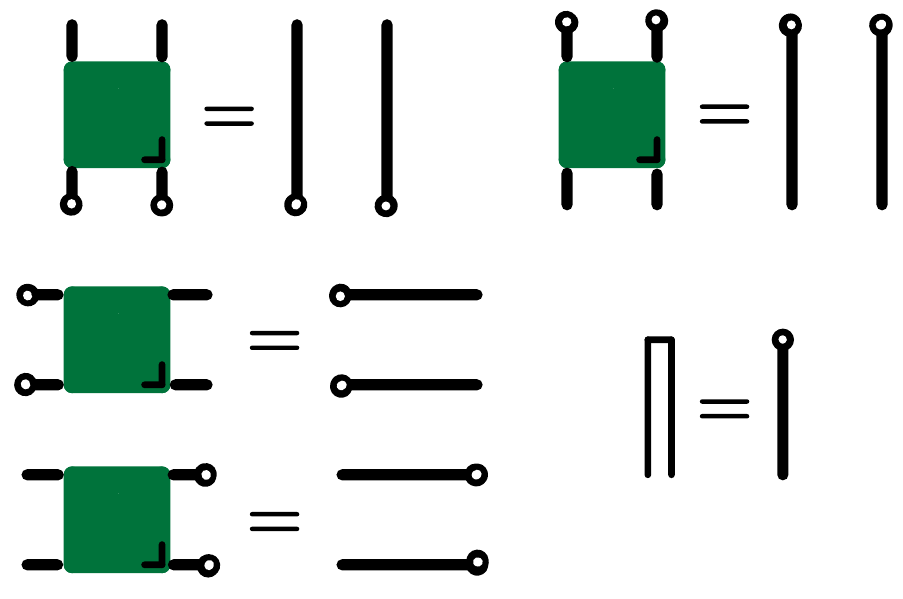}
\caption{\label{fig:6} 
Compact expressions of unitarity (top) and dual unitarity (bottom) for folded brick gates.}
\end{figure}

\begin{figure}
\includegraphics[scale=0.65]{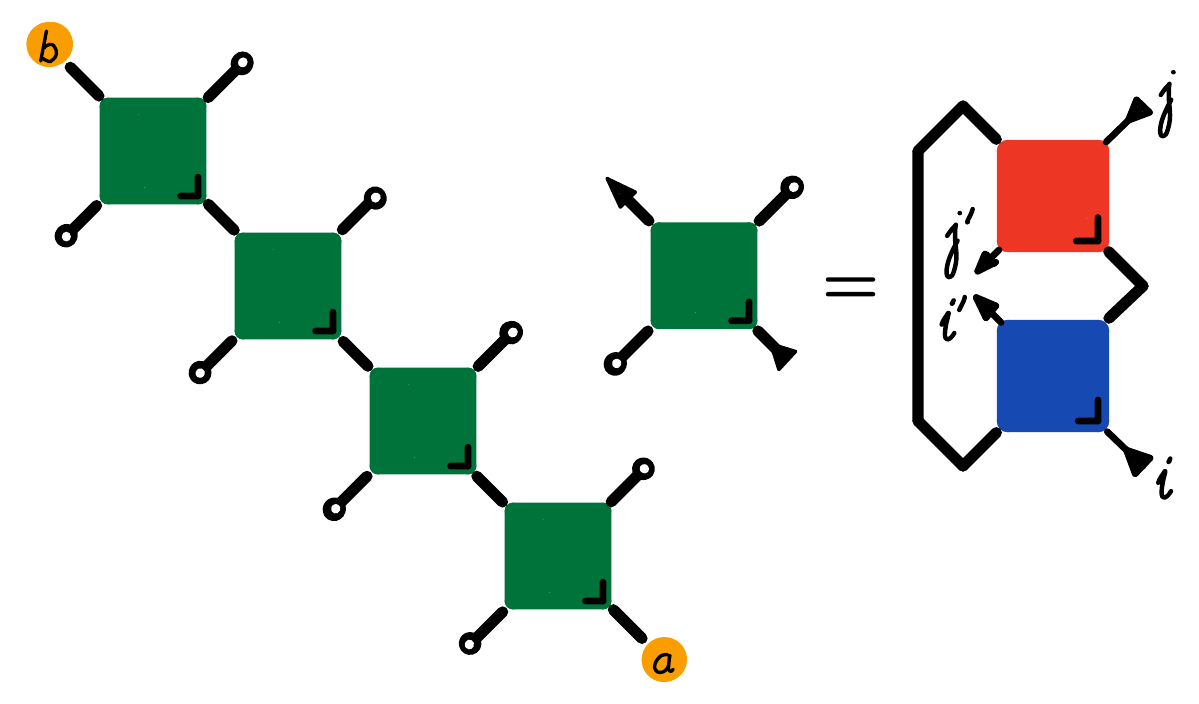}
\caption{\label{fig:8} 
The non-vanishing (light-ray) contribution to correlation function between local observables 
$a,b$ for the DU brickwork circuit -- using the folded circuit formulation -- (the second term of (\ref{eq:Cablr}) for $t=2$), and the definition of the corresponding transfer matrix $\mathcal M_-$ (right).}
\end{figure}

Computation of local spatiotemporal correlation functions of DU brickwork circuits can be largely simplified, namely it is easy to show that both, the (time) unitarity (\ref{eq:B1},\ref{eq:B2}), as well as the space unitarity (schematically depicted in Fig.~\ref{fig:6})
\begin{eqnarray}
\tilde{W}^\brick \kket{\circ}\otimes\kket{\circ} &=& \kket{\circ}\otimes\kket{\circ}, \label{eq:B3}\\
 \bbra{\circ}\otimes\bbra{\circ} \tilde{W}^\brick&=& \bbra{\circ}\otimes\bbra{\circ},\label{eq:B4}
 \end{eqnarray}
where $\tilde{W}^\brick = \tilde{U}^\brick\otimes (\tilde{U}^\brick)^T$ imply that the expression (\ref{eq:Cab}) vanishes unless $|x-y|=2t$. 
This is a consequence of causality within both, space-like and time-like cones, so the correlator can be non-vanishing only along two light-rays\cite{bertini2019exact}. There, it is expressed as
\begin{eqnarray}
C_{a,b}(x,y;t) &=& \delta_{y,x+2t}\delta_{{\rm mod}(x,2),1}
\,{\rm tr}\,(b \mathcal M_+^{2t}(a)) \nonumber\\
&+&  \delta_{y,x-2t}\delta_{{\rm mod}(x,2),0}
\,{\rm tr}\,(b \mathcal M_-^{2t}(a))
\label{eq:Cablr2}
\end{eqnarray}
in terms of completely positive, trace preserving and unital maps over ${\rm End}(\mathcal H_1)$, and 
${\rm End}(\mathcal H'_1)$, respectively (see Fig.~\ref{fig:8}),
\begin{eqnarray}
    \mathcal M_+ (a) &=& 
    \frac{1}{d'}({\rm tr}\otimes\One)\left( 
    (U^\brick)^\dagger 
    (a\otimes\one) U^\brick\right)\,,\\\
    \mathcal M_- (a) &=& 
    \frac{1}{d}(\One\otimes {\rm tr})\left( 
    (U^\brick)^\dagger 
    (\one\otimes a) U^\brick\right)\,.
\end{eqnarray}
$\One$ represents an identify map over the local space $\mathcal H_1^{(')}$, hence $\One\otimes{\rm tr}$ and ${\rm tr}\otimes\One$ denote the partial traces.
As $\mathcal M_\pm$ are linear non-expanding maps, their spectra are confined within the unit disk. Depending on whether there are additional eigenvalues, besides one eigenvalue $1$ corresponding to trivial eigenvector $\one$, which lie on the unit circle (respectively, at $1$), our Floquet circuit system is non-mixing (respectively, non-ergodic), otherwise it is mixing and ergodic. It has been shown in \cite{bertini2019exact} (and elaborated further in other DU models in \cite{austen_prl,arul21}) that one can have all different types of ergodic behavior even in the simplest class of DU brickwork circuits with $d=d'=2$. In the generic case (with probability 1 for a suitably random element of ${\rm DUBG}(d,d')$) the maps $\mathcal M_\pm$ have full rank ($d^2$ or ${d'}^2$) with all eigenvalues, except the trivial one, lying strictly inside the unit disk implying asymptotic \emph{exponential decay of correlations} (\ref{eq:Cab}) (mixing behavior) with the exponent given by the spectral gap of $\mathcal M_\pm$.

\subsection{Dual-unitary IRF circuits}

\begin{figure}
\includegraphics[scale=0.55]{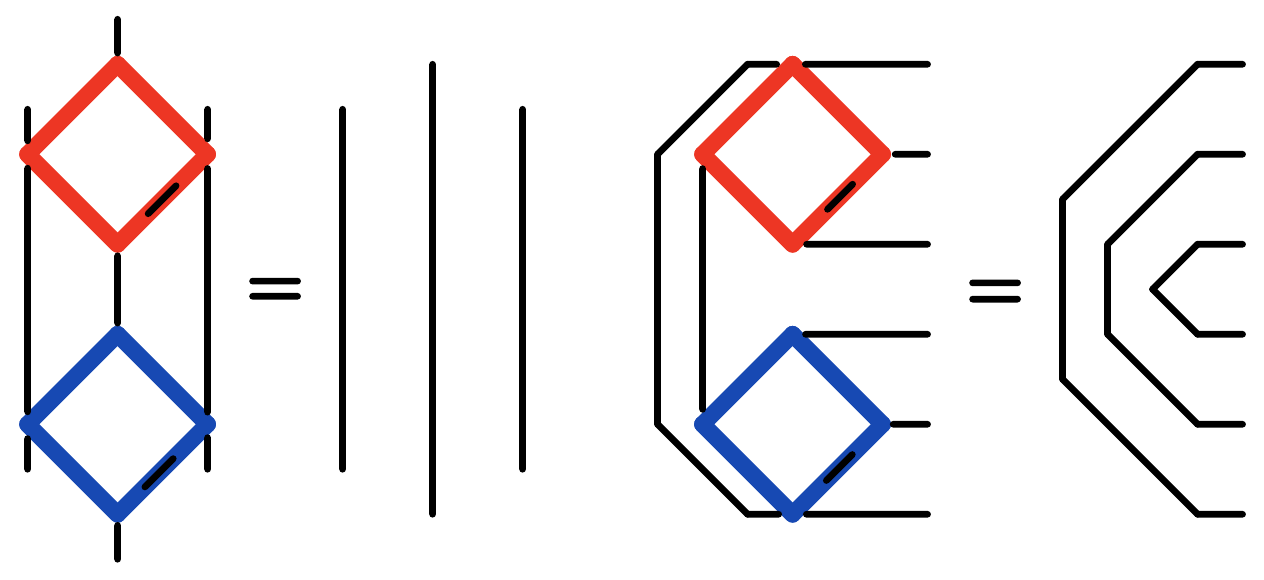}
\caption{\label{fig:4} 
Time unitarity (left) and space unitarity (right) condition for the dual-unitary IRF gate (element of ${\rm DUIRF}(d)$).}
\end{figure}

In somewhat close analogy to brickwork circuits we define DU IRF circuits, composed of IRF gate (\ref{eq:UIRF}), 
for which also the space-time dual $\tilde{U}^\IRF\in {\rm End}(\mathcal H_1^{\otimes 3})$:
\begin{equation}
    \tilde{U}^\IRF = \sum_{i,j,k,j'=1}^d (u_{jj'})_i^{k} \ket{i}\otimes \ket{j'}\otimes \ket{k} \bra{i}\otimes\bra{j}\otimes\bra{k}\,
    \label{eq:UtIRF}
\end{equation}
is unitary
\begin{equation}
    \tilde{U}^\IRF (\tilde{U}^\IRF)^\dagger = \one.
    \label{eq:IRFdu}
\end{equation}
This condition is equivalent to a condition that a set of $d^2$ (space-time flipped) matrices $\tilde{u}_{jj'} \in {\rm End}(\mathcal H_1)$, $j,j'=1,2\ldots,d$, defined as
\begin{equation}
    (\tilde{u}_{jj'})_{i}^k := (u_{ik})_{j}^{j'},
    \label{eq:ut}
\end{equation}
is unitary, $\tilde{u}_{jj'} \tilde{u}^\dagger_{jj'} = \one$, $j,j'=1,\ldots,d$. See Fig.~\ref{fig:4} for a diagrammatic illustration of these properties. 
In the next subsection \ref{d2IRF} we provide a complete parametrization of a set ${\rm DUIRF}(d)$ of DU IRF gates for $d=2$, while in section \ref{estimates} we estimate its dimensionality for larger $d$. 

\begin{figure}
\includegraphics[scale=0.6]{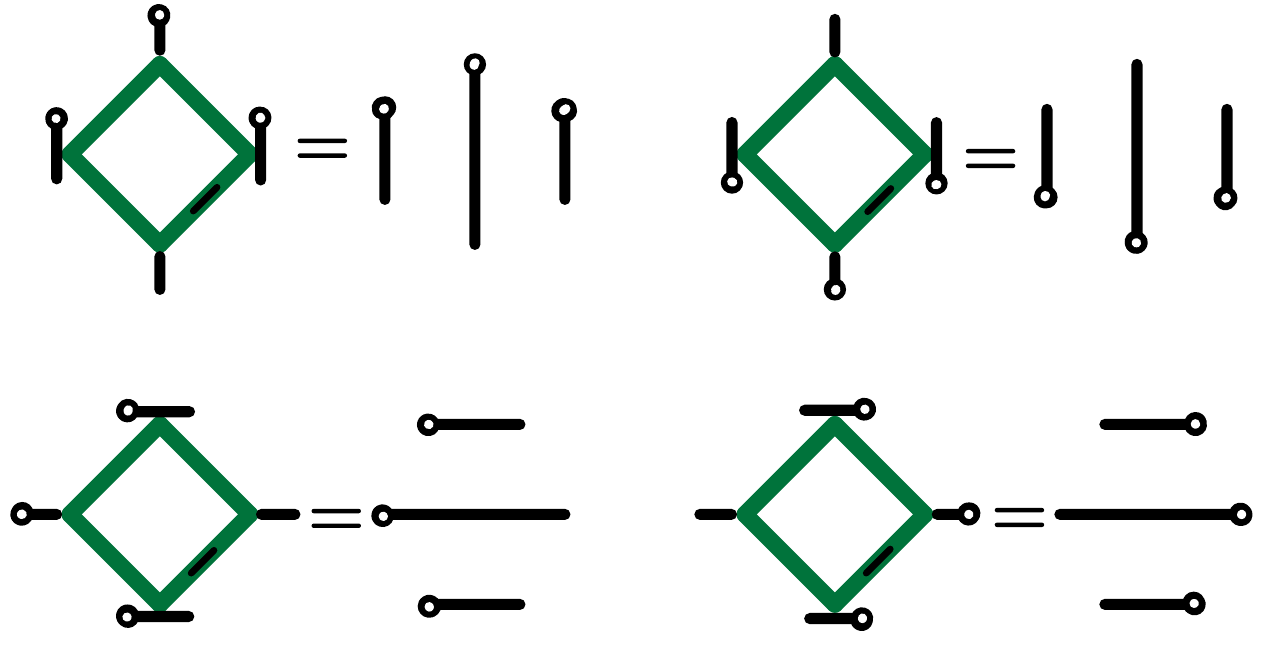}
\caption{\label{fig:7} 
Compact expressions of time unitarity (top) and space unitarity (bottom) for the folded IRF gates.}
\end{figure}

\begin{figure}
\includegraphics[scale=0.65]{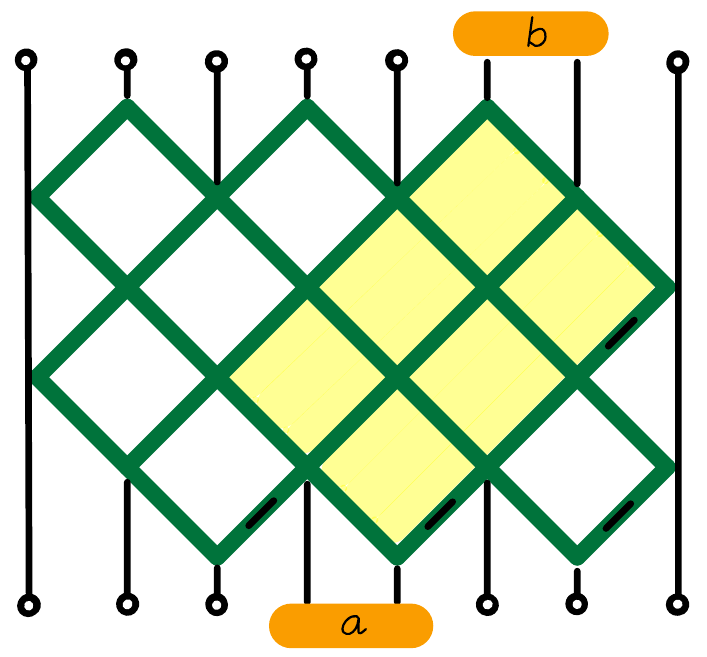}
\caption{\label{fig:9} 
Schematic illustration of computation of correlation function between local (2-site) observables in the folded IRF circuit formulation. The yellow-shaded area indicates the intersection of temporal causal cones to which the correlator can be simplified using only unitarity (Fig.~\ref{fig:7}-top). For DU IRF circuit one can apply (all) rules of Fig.~\ref{fig:7} to show that such correlation function identically vanishes (unless the supports of operators $a$ and $b$ are shifted precisely by $2t$, as used in Fig.~\ref{fig:10}.}
\end{figure}

\begin{figure}
\includegraphics[scale=0.65]{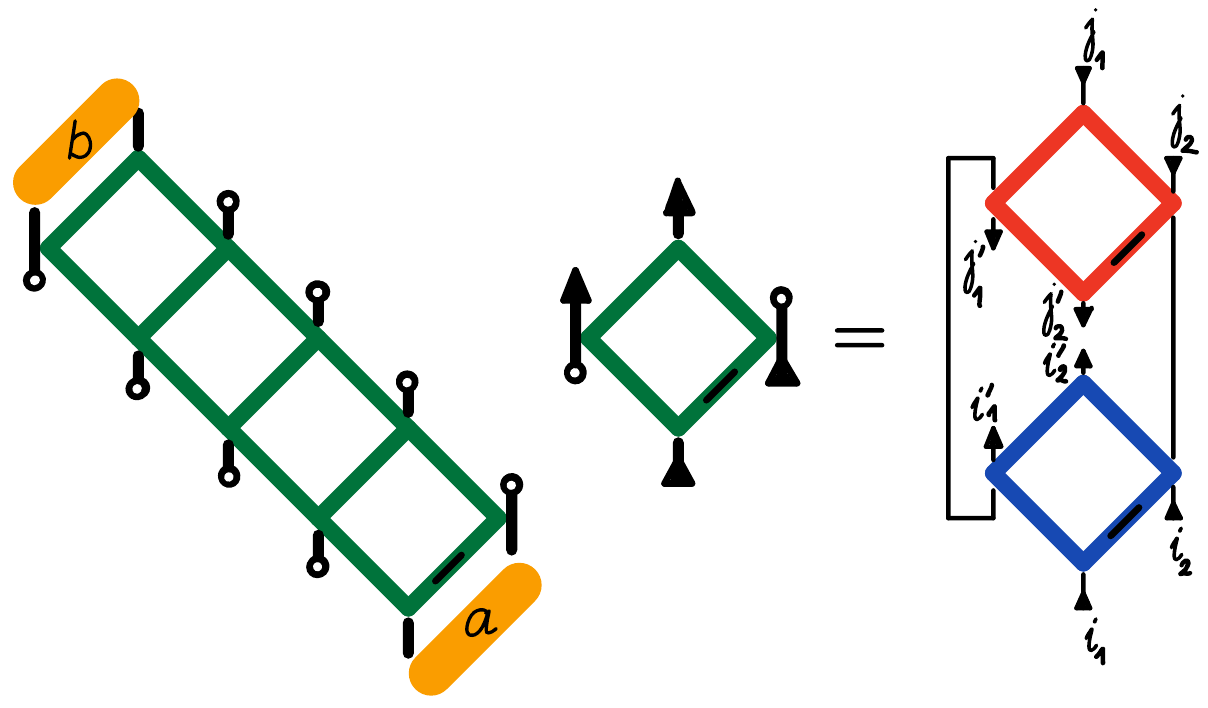}
\caption{\label{fig:10}
The nonvanishing (light-ray) contribution to correlation function between local observables 
$a,b$ for the DU IRF circuit -- using the folded circuit formulation -- and the definition of the corresponding transfer matrix $\mathcal K_-$ (right).
}
\end{figure}

In terms of the folded IRF gate $\tilde{W}^\IRF = \tilde{U}^\IRF\otimes(\tilde{U}^\IRF)^T$, cf. (\ref{eq:IRFfold}), the space unitarity (\ref{eq:IRFdu}) of DU IRF gate is elegantly expressed in terms of the second set of unitality conditions (graphically encoded in Fig.~\ref{fig:7}-bottom)
\begin{eqnarray}
\tilde{W}^\IRF \kket{\circ}\otimes\kket{\circ}
\otimes\kket{\circ}&=&\kket{\circ}\otimes\kket{\circ}\otimes\kket{\circ},\label{eq:I3}\\
 \bbra{\circ}\otimes\bbra{\circ}
\otimes\bbra{\circ}\tilde{W}^\IRF&=&\bbra{\circ}\otimes\bbra{\circ}\otimes\bbra{\circ}.\label{eq:I4}
\end{eqnarray}
The complete set of unitality relations (\ref{eq:I1},\ref{eq:I2},\ref{eq:I3},\ref{eq:I4}) is then facilitated to show that the correlator (\ref{eq:Cab}) (Fig.~\ref{fig:9}) vanishes unless $|x-y|=2t$. Without loss of generality we can now assume that local operators  are supported on two sites $a,b\in {\rm End}(\mathcal H_1^{\otimes 2})$ (including single-site observables which are trivial on the second site) and write $\kket{a_x} = \kket{\circ}^{\otimes (x-1)}\otimes\kket{a}\otimes\kket{\circ}^{\otimes(L-x-1)} $ and
$\bbra{b_y} =\bbra{\circ}^{\otimes(y-1)}\otimes\bbra{b}\otimes\bbra{\circ}^{\otimes(L-y-1)}$.

Diagrammatically, this is illustrated in Fig.~(\ref{fig:10}), where the resulting light-cone correlators:
\begin{eqnarray}
C_{a,b}(x,y;t) &=& \delta_{y,x+2t}\delta_{{\rm mod}(x,2),1}
\,{\rm tr}\,(b \mathcal K_+^{2t}(a)) \nonumber\\
&+&  \delta_{y,x-2t}\delta_{{\rm mod}(x,2),0}
\,{\rm tr}\,(b \mathcal K_-^{2t}(a))\,,
\label{eq:Cablr}
\end{eqnarray}
are expressed in terms of completely positive, trace preserving and unital maps over ${\rm End}(\mathcal H_1^{\otimes 2})$,
\begin{eqnarray}
    \mathcal K_+ (a) &=& 
    \frac{1}{d}({\rm tr}\otimes\One\otimes\One)\left( (U^\IRF)^\dagger 
    (a\otimes\one) U^\IRF\right)\,,\\\
    \mathcal K_- (a) &=& 
    \frac{1}{d}(\One\otimes \One\otimes {\rm tr})\left( (U^\IRF)^\dagger 
    (\one\otimes a) U^\IRF \right)\,,
\end{eqnarray}
(see Fig.~\ref{fig:10}-right for graphical defintion of $\mathcal K_-$).

Although the maps $\mathcal K_\pm$ act on a much larger (2-qudit) space as $\mathcal M_\pm$, they also have a large trivial subspace (of eigenvalue 0) and hence can be reduced to a simpler form. 
This essentially follows from the trivial action of the IRF gate on the control (left and right) qudits. Let 
\begin{equation}
\mathcal D ( \ket{j}\bra{j'}) = \delta_{j,j'} \ket{j}\bra{j'}
\end{equation}
represent a projector to diagonal subspace of ${\rm End}(\mathcal H_1)$.
The correlation maps clearly satisfy the identities (following from diagrammatics of Fig.~\ref{fig:10}):
\begin{eqnarray}
&&\mathcal K_+ (\mathcal D \otimes \One) = 
 (\One\otimes\mathcal D\mathcal) \mathcal K_+ = \mathcal K_+\,, \label{eq:KD}\\
&&\mathcal K_- (\One\otimes\mathcal D) = 
(\mathcal D \otimes \One) \mathcal K_- = \mathcal K_- \,. \nonumber
\end{eqnarray}
Defining the diagonally projected maps
\begin{equation}
    \mathcal K'_\pm = (\mathcal D\otimes \mathcal D)\mathcal K_\pm 
    (\mathcal D\otimes \mathcal D),\,
\end{equation}
and using the projector property $\mathcal D^2=\mathcal D$,
one finds that Eqs. (\ref{eq:KD}) imply, for any $t\in\mathbb Z$:
\begin{equation}
 (\mathcal D\otimes \mathcal D)(\mathcal K_\pm)^t (\mathcal D\otimes \mathcal D)
 = (\mathcal K'_\pm)^t\,.
\end{equation}
This in turn implies that the correlation functions (\ref{eq:Cablr2}) are  given in terms of simple iteration of diagonally projected maps
\begin{equation}
    {\rm tr}\left(b \mathcal K^t_\pm (a)\right) = 
    {\rm tr}\left(b_{\rm d}(\mathcal K'_\pm)^t (a_{\rm d})\right)
\end{equation}
where $a_{\rm d} = \mathcal D\otimes\mathcal D a$, $b_{\rm d} = \mathcal D\otimes\mathcal D b$ are diagonal (projected) 2-site observables.
In fact the maps $\mathcal K'_\pm$ can be identified with the classical Markov chains. By identifying the basis $\{j \leftarrow \ket{j}\bra{j}\}$, the explicit matrix representation of correlation maps reads
\begin{equation}
    (\mathcal K'_+)_{i\,j}^{i'j'} = \frac{1}{d}\left| (u_{ij'})_j^{i'} \right|^2\,,\quad
    (\mathcal K'_-)_{i\,j}^{i'j'} = \frac{1}{d}\left| (u_{i'j})_i^{j'} \right|^2\,.
    \label{Kmarkov}
\end{equation}
These matrices are bistochastic. In fact, they are bistochastic also under the flip of indices
($j\leftrightarrow i'$) which would correspond to space-time flip if one composes from them a brickwork classical Markov circuit like those studied in Ref.~\cite{PRX2021}, hence they may be referred to as {\em dual bistochastic}.\footnote{Note, however, that in spite of some formal similarities these matrices are not unistochastic.}

It follows from the form (\ref{Kmarkov}) and unitarity of 
$u_{ik}$ and $\tilde{u}_{jj'}$ that the map $\mathcal K'_\pm$ annihilates the diagonal operators of the form $\one\otimes a_{\rm d}$ or $a_{\rm d} \otimes \one$, where $a_{\rm d}\in {\rm End}(\mathcal H_1)$, ${\rm tr}\, a_{\rm d}=0$.
Hence $\mathcal K'_\pm$ act nontrivially within a subspace spanned by $\one$ and traceless operators supported on no less than 2 neighbouring sites, which yields their maximal rank
\begin{equation}
    \max\,{\rm rank}\,\mathcal K_\pm = 1 + (d-1)^2.
    \label{maxrank}
\end{equation}

The above observation also implies that all correlation functions between single-site (ultra-local) observables vanish, 
while the simplest non-trivial correlations involve two-site observables. In summary, the decay of correlation functions of local observables in DU IRF circuits is thus completely determined by the spectra of dual bistochastic $d^2\times d^2$ matrices $\mathcal K'_\pm$ (in fact, by their $(d-1)^2$ dimensional nontrivial blocks) and the absence of nontrivial eigenvalue $1$ (respectively, unimodular eigenvalue) signals ergodic (respectively, mixing) dynamics. 

\subsection{Complete parametrization of dual-unitary IRF qubit gates}
\label{d2IRF}

Let us now consider the case $d=2$ with an attempt to parametrize all DU IRF gates. We start by Euler angle parametrization of
${\rm U}(2)$ matrices $u_{ik}$
\begin{equation}
    u_{ik} = e^{{\rm i} \phi_{ik}} 
    \begin{pmatrix}
    e^{{\rm i}\nu_{ik}}\cos\theta_{ik} &
    e^{{\rm i}\eta_{ik}}\sin\theta_{ik} \cr
    -e^{-{\rm i}\eta_{ik}}\sin\theta_{ik} &
    e^{-{\rm i}\nu_{ik}}\cos\theta_{ik} 
    \end{pmatrix}\,,
\end{equation}
where $\phi_{ik},\nu_{ik},\eta_{ik},\theta_{ik}\in [0,2\pi)$, $i,k=1,2$, are 16 real parameters (note that such parametrization is non-injective).
Solving for unitarity of $\tilde{u}_{jj'}$, defined in (\ref{eq:ut}), separates nicely into two sets of equations: The equations for $\theta_{ik}$
\begin{eqnarray}
    &\cos^2\theta_{11}=\sin^2\theta_{12},\quad
    \cos^2\theta_{22}=\sin^2\theta_{21}, \nonumber\\
    &\cos\theta_{11}\cos\theta_{21}+\cos\theta_{12}\cos\theta_{22}=0, \label{eqtheta}\\ &\sin\theta_{11}\sin\theta_{21}+\sin\theta_{12}\sin\theta_{22}=0, \nonumber
\end{eqnarray}
and a set of linear equations for the other variables which fixes, say $22$-components of the angles $\nu_{ik},\eta_{ik},\phi_{ik}$ in terms of components $11,12,21$:
\begin{eqnarray}
\nu_{22}&=&\nu_{12}+\nu_{21}-\nu_{11},\nonumber\\
\eta_{22}&=&\eta_{12}+\eta_{21}-\eta_{11},\\
\phi_{22}&=&\phi_{12}+\phi_{21}-\phi_{11}.\nonumber 
\end{eqnarray}
Eqs.~(\ref{eqtheta}) in turn result in expressing 
three $\theta_{ik}$ in terms of the fourth, say $\theta_{22}$. There are two equivalent solutions, while without loss of generality we take:
\begin{equation}
\theta_{11} = \theta_{22} + \pi,\quad
\theta_{12} = \theta_{21} = \theta_{22} + \frac{\pi}{2}.
\end{equation}
We thus parametrized ${\rm DUIRF}(2)$ in terms of 10 independent free parameters 
$\{ \theta_{22},\nu_{11},\nu_{12},\nu_{21},\eta_{11},\eta_{12},\eta_{21},\phi_{11},\phi_{12},\phi_{21}\}$, hence ${\rm dim}\,{\rm DUIRF}(2)={\bf 10}$. Considering 4-dimensional gauge symmetry (\ref{gaugeIRF}) and a global (overall) phase, we have in fact $10-4-1={\bf 5}$ parametric set of physically inequivalent IRF gates of qubits.

Expressing the diagonally projected transfer matrices we obtain a simple result
\begin{equation}
    \mathcal K'_+ = \mathcal K'_- = 
    \frac{1}{2}\begin{pmatrix}
    \cos^2\theta_{22} & \sin^2\theta_{22} & \sin^2\theta_{22} & \cos^2\theta_{22} \\
     \sin^2\theta_{22} & \cos^2\theta_{22} & \cos^2\theta_{22} & \sin^2\theta_{22} \\
      \sin^2\theta_{22} & \cos^2\theta_{22} & \cos^2\theta_{22} & \sin^2\theta_{22} \\
      \cos^2\theta_{22} & \sin^2\theta_{22} & \sin^2\theta_{22} & \cos^2\theta_{22} 
    \end{pmatrix}.
\end{equation}
$\mathcal K'_\pm$ have rank 2 and a single nontrivial eigenvalue $\lambda = \cos(2\theta_{22})$ with the corresponding left\&right eigenvector $(1,-1,-1,1)$ corresponding to eigenoperator $a = \sigma\otimes\sigma$.
We have thus shown that the only nontrivial (nonzero) correlation function -- autocorrelation of the Ising interaction -- of all DU IRF circuits with $d=2$ has a universal form
\begin{equation}
    C_{\sigma\otimes\sigma,\sigma\otimes\sigma}(x,y;t)  
     = (\delta_{y,x+2t}\delta_{{\rm mod}(x,2),1}+ \delta_{y,x-2t}\delta_{{\rm mod}(x,2),0}) \lambda^{2t},
\end{equation}
independent of all other parameters (but $\theta_{22}$) of the gate.\footnote{Note that this feature is similar to a kicked Hadamard spin chains studied in Ref.\cite{Gutkin_Hadamard}. It would be interesting to investigate if theres is a link between those sistems and DU IRF circuits.}

Of course, we expect, and find, that DU IRF curcits for higher $d$ have much richer behavior, as reported in the next section.

\section{Estimating the dimension of manifolds of Dual-unitary gates for $d> 2$ ($d'>2$)}
\label{estimates}

This is an experimental section of the paper where we provide some empirical observation which can hopefully guide further progress. Earlier we managed to fully characterize the manifolds of DU brick and IRF gates for qubits $d=2$. It has become clear that obtaining rigorous results in this direction for DU brick gates with $d>2$ or $d'>2$ is notoriously difficult, while this task does not appear to get any easier for DU IRF gates with $d>2$.

Therefore we take a different approach here and try to estimate numerically the number of free real parameters (dimensionality) of ${\rm DUBG}(d,d')$ and ${\rm DUIRF}(d)$.
We do this by determining the dimensions of tangent spaces at {\em random} instances of solutions to dual (time and space) unitarity conditions.

\begin{table}
\label{table}
\begin{ruledtabular}
\begin{tabular}{ccccc}
$d$ & $d'$ & $N=(d d')^2$ &
${\rm dim}\,{\rm DUBG}(d,d')$ &
${\rm dim}\,{\rm DUIRF}(d)$ \\
\hline
2 & 2 & 16 & 12 (12) & 10 (10) \\
3 & 3 & 81 & 45 (45,43,41) & (33,29,25) \\
4 & 4 & 256 & 112 (94) & (49) \\
5 & 5 & 625 & 225 (97) & (81) \\
6 & 6 & 1296 & 396 (141) & (121) \\
7 & 7 & 2401 & 637 (193) & (169) \\
3 & 2 & 36 & 24 (24) & --- \\
4 & 2 & 64 & 40 (40) & --- \\
5 & 2 & 100 & 60 (60) & --- \\
6 & 2 & 144 & 84 (84) & --- \\
7 & 2 & 196 & 112 (112) & --- \\
4 & 3 & 144 & 72 (66,64) & --- \\
5 & 3 & 225 & 105 (77) & --- \\
\end{tabular}
\end{ruledtabular}
\caption{Local dimensions of manifolds ${\rm DUBG}(d,d')$ and ${\rm DUIRF}(d)$ estimated as dimensions of the tangent spaces at randomly sampled solutions of DU constraints (numbers within brackets). For comparison we show for {\rm DUBG} also dimensions of tangent spaces at random instances of explicit parametrization \cite{balazs_private} of DU brick gates of Eq.~(\ref{balazs}) (unbracketed numbers).}
\end{table}

\subsection{Dual unitary brick gate manifolds}

Writing $N=2(d d')^2$ real components of a $dd' \times dd'$ complex matrix $ U^\brick $ in terms of a vector $\vec{z}=(z_1,z_2,\ldots,z_{N})$ we can write the dual unitarity conditions
$ U^\brick (U^\brick)^\dagger = \one$, $\tilde{U}^\brick (\tilde{U}^\brick)^\dagger=\one$, in terms of a zero of a nonlinear (quadratic) vector function $\vec{f}(\vec{z})$. Note that the number $M$ of components of $\vec{f}$ (number of equations) is in general different (larger) than the number of variables $N$.

Considering an instance $\vec{z}_*$ of a solution $\vec{f}(\vec{z}_*) = \vec{0}$, corresponding to an elelement $U^\brick \in {\rm DUBG}(d,d')$ we can estimate a \emph{local dimension} ${\rm dim}(\vec{z}_*)$ of ${\rm DUBG}(d,d')$ as the dimension of the tangent space, i.e. by the rank of the $M\times N$ derivative matrix
$F(\vec{z}_*) = \{\partial f_i (\vec{z}_*)/\partial z_j\}^{i=1\ldots M}_{j=1\ldots N}$,
which is numerically determined by the number of nonvanishing singular values of $F(\vec{z}_*)$:
\begin{equation}
    {\rm dim}(\vec{z}_*) = N - {\rm rank}\, F(\vec{z}_*)\,.
    \label{dim}
\end{equation}
If ${\rm DUBG}(d,d')$ were a simple manifold the dimension should not depend on the point 
$\vec{z}_*\in {\rm DUBG}(d,d')$. This, however, does not seem to be the case when both $d,d'\ge 3$, so different pieces of the set ${\rm DUBG}(d,d')$ may have different topological dimensions.

We made the following numerical experiment. We sampled an ensemble of the order of $10^2 - 10^4$ (depending on values of $d,d'$) random solutions $\vec{z}_{*}$ of $\vec{f}(\vec{z}_*)=\vec{0}$ which were obtained by
running Wolfram's Mathematica routine {\tt FindMinimum} on $|\vec{f}(\vec{z})|^2$ applied to random initial seeds where $z_j$ were i.i.d. Gaussian random with zero mean and variance $1/(dd')$ (reproducing unitarity in the limit $d,d'\to\infty$). We note that random instances of DU gates could also be generated by iteration of a non-linear map proposed in Ref.~\cite{arul_prl2020}.
We have then determined the possible values of
${\rm dim}(\vec{z}_*)$ and collected them in 
Table I. 
We note that numerical values of $|\vec{f}(\vec{z}_*)|$ were typically between $10^{-13}$ and $10^{-7}$ and there was always a clear cutoff in the singular value spectrum of $F(\vec{z}^*)$, where the `zero' singular values were at least five orders of magnitude smaller than the rest.
 For $d=d'=2$ we reproduce the expected analytical result ${\rm dim}=12$. For $d=d'=3$ we obtain three different values ${\rm dim}=45,43,41$ within our data, while for larger $d,d'$ we empirically find only a single dimension (as shown within brackets in Table I). We suspect that for $d,d'>3$ we do not observe other (higher) local dimensions simply because hitting such solutions $\vec{z}_*$ becomes statistically increasingly unlikely. We confirm this speculation by analysing the parametrization that has been proposed by Bal\' azs Pozsgay \cite{balazs_private}.
 Specifically, one can write the elements of essentially the largest known subset of ${\rm DUBG}(d,d')$ as 
 \begin{equation}
U^\brick = S\sum_{s=1}^{d'} u_s \otimes \ket{s}\bra{s},
\label{balazs}
 \end{equation}
where $S\ket{j}\otimes\ket{s}=\ket{s}\otimes\ket{j}$.
 Clearly, $U^\brick$ is dual-unitary when $u_s$ are arbitrary $d\times d $ unitary matrices. This gives us $d^2 d'$ free parameters, where we can assume $d\ge d'$ without loss of generality. It is striking that empirically found local dimensions in Table I are smaller than $d^3$ for $d=d'\ge 4$. However, determining the dimension (\ref{dim}) for $\vec{z}_*$ parametrizing random DU brick gate of the form (\ref{balazs}) (considering $u_s$ as Haar random) we obtain consistent (non-fluctuating) numbers considerably larger than $d^2 d'$ shown as unbracketed numbers in Table I.
 
 When the smaller of the spaces is a qubit ($d'=2$) we consistenly find that all random solutions of $\vec{f}(\vec{z}_*)=\vec{0}$ give the same dimension (as well as random instances of (\ref{balazs})), so the complexity/topology of the space 
 ${\rm DUBG}(d,d')$ seems fundamentally different when both spaces are non-qubit.
 
 We have also checked that the correlation maps $\mathcal M_\pm$ have full ranks, $d^2$ and ${d'}^2$ respectively, for all randomly generated solution instances $\vec{z}_*$.
\\\\
 {\bf\em Conjecture:} Data of Table I suggest a clear conjecture on dimensions of the manifolds of dual unitary brick gates: (i) When one of the spaces is a qubit ($d'=2$) we find a simple quadratic scaling of manifold dimensions
 \begin{equation}
 {\rm dim}\,{\rm DUBG}(d,2) = 2 (d+1)d\,,
 \end{equation}
while (ii) in general we find that, although the local dimensions are fluctuating, the maximal dimension (tangent to (\ref{balazs}) is perfectly fitted by a cubic polynomial in $d,d'$
\begin{eqnarray}
    &&{\rm max}\,{\rm dim}\,{\rm DUBG}(d,d') = \\ 
    &&(d^2+{d'}^2)d' +(4d-5d')d' + 6(d'-d)\,,\quad{\rm if}\;\; d\ge d'\,,
    \nonumber
\end{eqnarray}
with $d,d'$ swapped if $d\le d'$. For $d=d'$ we have 
${\rm max}\,{\rm dim}= (2d-1)d^2$.

\subsection{Dual unitary IRF manifolds}

Completely analogous Mathematica program has been developed for targeting the local dimensions of the manifold ${\rm DUIRF}(d)$ where $\vec{f}(\vec{z})$ now encodes the constraints on dual (space and time) unitarity of the IRF gate, while vector $\vec{z}$ completely encodes the matrices $u_{ik}$, and initial seed variables $z_n$ were taken as Gaussian i.i.d. with zero mean and variance $1/d$. We have found consistently smaller dimensions than for ${\rm DUBG}(d,d)$ (see Table I). We reproduced the correct analytic result ${\rm dim}=10$ for $d=2$, and again, for $d=3$, we have found multiple local dimensions depending on the instance of the solution $\vec{f}(\vec{z}_*)=0$. For $d\ge 4$ the empirical dimensions were again unique, but this might be a statistical effect, like in the case of DU brick gates. Obtaining a systematic (analytic) parametrization (of large subsets) of ${\rm DUIRF}(d)$ for $d\ge 3$ remains an open problem.

We have checked as well that the correlation maps $\mathcal K'_\pm$ have maximal ranks (\ref{maxrank}), $d^2-2(d-1)$, for all randomly generated solution instances $\vec{z}_*$.

\section{Discussion}

Dual unitary circuits, either in brickwork or IRF form, allow for an exact reduction of dynamics of an interacting theory in 1+1 (space-time) dimensions to an open, dissipative (markovian) quantum dynamics of a single particle (qudit) or a pair of particles (qudits). Although this connection has so far been elaborated only for ultra-local (for brickwork circuits) or $2-$local (for IRF circuits) 
observables, it is straightforward to generalise it to observables with arbitrary finite range support: the resulting quantum Markov chain process is then defined on a sufficiently large (finite) set of qudits reflecting the range of the observables.

We expect that other recent exact results on dynamics of DU brickwork circuits can be extended to DU IRF circits. Firstly, one may attempt to compute the \emph{spectral form factor} written as\cite{BKP20}
\begin{equation}
    K(t) = \mathbb E\left( 
    \left|{\rm tr}\, \mathcal U^t \right|^2\right)
    = \mathbb E\left( {\rm tr}\, \mathcal W^t\right)
\end{equation}
where $\mathbb E$ represents a suitable quench-disorder averaging.
The natural form of disorder is now a random unitary diagonal transformation $\Delta_x={\rm diag}\{ e^{{\rm i}h^{(j)}_x}; j=1,\ldots,d\}$ at each site $x$, after each half-time-step (\ref{eq:UF}), which preserves the DU IRF gate structure, where the fields $h^{(j)}_x$ are i.i.d. random variables (of essentially arbitrary smooth distribution). After a space-time flip we can perform averaging locally (in full analogy to Refs.~\cite{BKP18,BKP20}) and write
\begin{equation}
   K(t) = \mathbb E\left( {\rm tr}\, {\mathcal{\tilde{W}}}^L\right) =  
   {\rm tr} \left(\mathbb E \mathcal{\tilde{W}}\right)^L,\,
\end{equation}
where $\tilde{\mathcal W}$ denotes the folded circuit propagator in space direction, composed as Eqs. (\ref{eq:UF},\ref{eq:IRFC}) with $U^\IRF$ replaced by $\tilde{W}^\IRF$ and $L$ replaced by $t$ (which depends on site-independent random fields $h^{(j)}_x$ at fixed $x$).
Averaging should then again result\cite{BKP18,BKP20} in additional non-expanding piece $\mathcal O$ to the transfer matrix
$\mathcal T = 
\mathbb E \tilde{\mathcal{W}} = \mathcal O \tilde{\mathcal{W}}.
$
Translational invariance in time (together with the structure of the generic gate $U^{\rm IRF}$ and absence of time-reversal symmetry) should then result in $\mathcal T$ having exactly $t$-dimensional eigenspace of eigenvalue $1$ and the rest of the spectrum gapped within the unit disk, resulting in exact RMT expression in the thermodynamic limit $\lim_{L\to\infty} K(t) = t$.

Similarly, we expect that the explicit results on solvable initial states in quenched DU circuits \cite{Lorenzo}, as well as on maximal growth rate of entanglement\cite{PRX19}, operator spreading \cite{bertini2020operatorI,bertini2020operatorII,Reid}, and tripartite information \cite{BrunoLorenzo}, should have their close analogs for DU IRF circuits. 

Lastly, it is an interesting open question if one can find mappings between brickwork and IRF circuit models on abstract level, similarly as the class of integrable IRF models could be understood in terms of a subalgebra of integrable vertex models\cite{pasquier}.


\begin{acknowledgments}
The author warmly acknowledges Bruno Bertini and Pavel Kos for very fruitful collaboration on closely related topics, and to Bal\' azs Pozsgay and Vedika Khemani for inspiring discussions.
In particular, Bal\' azs Pozsgay is acknowledged for providing the
parametrization (\ref{balazs}). 
This work has been supported by the European Research Council (ERC) under the
Advanced Grant No.\ 694544 -- OMNES, and by the Slovenian Research Agency (ARRS)
under the Program P1-0402.
\end{acknowledgments}


\bibliography{bibliography}

\end{document}